\documentclass[twocolumn,10pt]{asme2e}
					\usepackage{float}
						\usepackage[cmex10]{amsmath}
						\usepackage{newtxtext,newtxmath}	
						\usepackage{bm}
					\usepackage{siunitx}
						\usepackage{graphicx}
					    \usepackage{epsfig,epstopdf}
					
				    \usepackage[font={footnotesize,sf,bf}]{caption}
			        \usepackage[font={footnotesize,sf,bf}]{subcaption}

						\usepackage{tabu}
						\usepackage{multirow}
						\usepackage{hhline}
						\usepackage{booktabs}
						\usepackage{setspace}
					
						\usepackage{cite}
					
						\usepackage{nomencl}
							\makenomenclature
						\usepackage[colorlinks=true,
									linkcolor=blue,
									citecolor=blue,
									urlcolor=blue,
									filecolor=blue,
									 pdfpagemode=None,
									pdfstartview=FitH]{hyperref}
					            
						\usepackage{enumerate}
					
					\usepackage{tikz,pgfplots}
					\usetikzlibrary{positioning, arrows.meta}
					\tikzset{%
						myarrow/.style = {-Stealth, shorten >=5pt}
					}
						\newcommand{\mvect}[1]{\mathbf{#1}}
						
						\newcommand{\trr}[1]{{#1}^{\!\top}}
						
						\newcommand{\inv}[1]{{#1}^{\text{-}1}}	
					\usepackage{algorithm}
					\usepackage{algorithmic}
					
					\usepackage{titlesec}
					\titlespacing{\section}{0pt}{1\baselineskip}{0pt}
					\titlespacing{\subsection}{0pt}{1\baselineskip}{0pt}			
					\newenvironment{rcases}
					{\left.\begin{aligned}}
						{\end{aligned}\right\rbrace}
	\newcommand{\PKadd}[1]{\textcolor{black}{#1}}

					\title{Topology optimization of fluidic pressure-driven multi-material compliant mechanisms}
					
					\author{Prabhat Kumar
								\affiliation{
								Mechanical and Aerospace Engineering,\\
								IIT Hyderabad,
								Telangana, 502285, India \\
								Email: pkumar@mae.iith.ac.in			
								}
					}
					\author{Josh Pinskier
							\affiliation{
								Robotics and Autonomous Systems Group,\\ CSIRO Data61,
								Brisbane, Australia \\
								Email: josh.pinskier@csiro.au
							}
					}
					\author{David Howard
							\affiliation{
								Robotics and Autonomous Systems Group, \\CSIRO Data61,
								Brisbane, Australia\\
								Email: david.howard@csiro.au
							}
					}
								\author{Matthijs Langelaar
					\affiliation{
						Department of Precision and Microsystems Engineering, \\TU Delft,
						The Netherlands \\
						Email: m.langelaar@tudelft.nl 
					}
				}
					
\begin{document}
					\small
					\maketitle
					\begin{abstract}
					\it{Compliant mechanisms actuated by pneumatic loads are receiving increasing attention due to their direct applicability as soft robots that perform tasks using their flexible bodies. Using multiple materials to build them can further improve their performance and efficiency. Due to developments in additive manufacturing, the fabrication of multi-material soft robots is becoming a real possibility. To exploit this opportunity, there is a need for a dedicated design approach. This paper offers a systematic approach to developing such mechanisms using topology optimization. The extended SIMP scheme is employed for multi-material modeling. The design-dependent nature of the pressure load is modeled using the Darcy law with a volumetric drainage term. Flow coefficient of each element is interpolated using a smoothed Heaviside function. The obtained pressure field is converted to consistent nodal loads. The adjoint-variable approach is employed to determine the sensitivities. A robust formulation is employed, wherein a min-max optimization problem is formulated using the output displacements of the eroded and blueprint designs. Volume constraints are applied to the blueprint design, whereas the strain energy constraint is formulated with respect to the eroded design. The efficacy and success of the approach are demonstrated by designing pneumatically actuated multi-material gripper and contractor mechanisms. A numerical study confirms that multiple-material mechanisms perform relatively better than their single-material counterparts.}
					
					\end{abstract}
					
					\section{Introduction} \label{I}
Compliant mechanisms (CMs) perform complex tasks by using the deformation of their own body. Typically, they are actuated via constant (design-independent) input forces/loads~\cite{howell2001compliant,zhu2020design}. Nowadays, demands for pneumatically (design-dependent loads) actuated compliant mechanisms as soft robots typically made from flexible rubber-like materials for various applications, e.g., soft gripping, force/displacement inverters/magnifiers, etc., is constantly increasing in industry and academia. Such mechanisms provide various advantages~\cite{xavier2022soft,kumar2023towards}. The performances of these mechanisms can further be enhanced when they are built using multiple materials. In addition, due to developments in additive manufacturing, the fabrication of multi-material soft robots is becoming a real possibility\cite{bandyopadhyay2018additive,langelaar2016topology}. To exploit this opportunity, there is a need for a dedicated systematic design approach. The primary aim of this paper is to present a systematic approach to design such multi-material pneumatically actuated mechanisms using topology optimization. 
		
				
		Topology optimization (TO) is a fast-growing design technique~\cite{sigmund2013topology,van2013level}. It enables the introduction and removal of solid and void regions during the structure evolution while extremizing an objective under the given physical/material constraints. In a typical structural setting, the design domain is parameterized by a set of finite elements. Each element is assigned a density/design variable~$\rho \in [0,\,1]$, which is assumed to be constant within the element. $\rho_i= 1$ and $\rho_i =0$ indicate element~$i$ is filled with material (solid-state) and no material (void phase), respectively. Ideally, the optimized designs consist of elements with $\rho=1$. However, one gets fictitious material within some elements, i.e., elements with $0<\rho<1$ exist in the optimized design, as the TO problem is relaxed to achieve a  solution~\cite{sigmund2013topology}. Considering more than one candidate material within a TO setting provides a much larger design space for the optimizer to explore; thus, one can obtain optimized designs with superior performances and high efficiency. 
			
			 \begin{figure*}[h!]
				\centering
				\begin{subfigure}[t]{0.450\textwidth}
					\centering
					\includegraphics[scale=1]{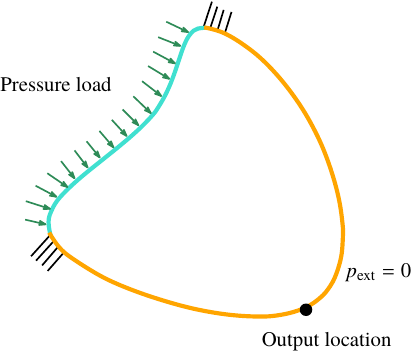}
					\caption{}
					\label{fig:Schematic1}
				\end{subfigure}
	\hfill
				\begin{subfigure}[t]{0.45\textwidth}
					\centering
					\includegraphics[scale=1]{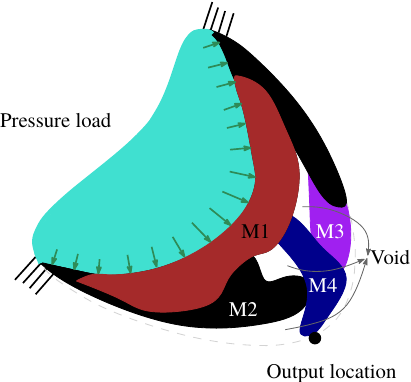}
					\caption{}
					\label{fig:Schematic2}
				\end{subfigure}
				\caption{A schematic diagram for a pressure-driven multi-material compliant mechanism. (\subref{fig:Schematic1}) Design domain. Pressure load is indicated via a set of arrows. Fixed boundary conditions are also depicted.  (\subref{fig:Schematic2}) A representative solution. $M_i|{_{i = 1,\,2,3,\,4}}$ are the candidate materials. One notices that with design changes the pressure load moves to a new position; thus, its magnitude, direction, and location get altered. Note here we depict only the initial and final pressure loading surfaces in (\subref{fig:Schematic1}) and (\subref{fig:Schematic2}), respectively.} \label{fig:Schematic}
			\end{figure*}   
		
Thomsen~\cite{thomsen1992topology} was the first to propose a multi-material TO approach by considering the concentration and orientation of composites' fibers as design variables. Sigmund and Torquato~\cite{sigmund1997design} presented a multi-material TO method by extending the SIMP (Solid Isotropic Material with Penalization) formulation for designing materials with extreme thermal expansion. The number of design variables per element is the same as the candidate materials; thus, the size of the optimization problem increases with the number of materials. This scheme is also termed recursive multiphase material interpolation in~\cite{gao2011mass}. Fujii et al.~\cite{fujii2001composite} proposed the homogenization-based multi-material method. The peak function was presented by Yin and Anathasuresh~\cite{yin2001topology} for designing multi-material CMs. The technique uses only one design variable per element; however, it requires gradual parameter tuning to achieve different peaks. Chu et al.~\cite{chu2018stress} presented a stress-based multi-material topology optimization for CMs. Gaynor\cite{gaynor2014multiple} used the extended SIMP material modeling with the robust formulation for designing multi-material CMs. An ordered SIMP formulation was presented by Zuo and Saitou~\cite{zuo2017multi}. Sivapuram et al.~\cite{sivapuram2021design} gave a TO method with many materials using integer programming and the extended SIMP approach. We also employ the extended SIMP formulation for multi-material modeling~\cite{sigmund1997design,sivapuram2021design} with the design-dependent pneumatic load.
				
Fluidic pressure loads show design-dependent behaviors as they alter the location, magnitude, and direction with the structural evolution during TO. Fig.~\ref{fig:Schematic1} shows a schematic multi-material mechanism problem pneumatic load. A representative solution is depicted in Fig.~\ref{fig:Schematic2}. One notices that the pressure load changes location and direction as the loading surface gets altered in the final topology. Therefore, modeling such design-dependent loads becomes challenging within a TO setting~\cite{hammer2000topology,kumar2020topology,sigmund2007topology,picelli2019topology,kumar2023TOPress}.

As pneumatically actuated mechanisms experience finite deformation, which poses several unique challenges within a TO setting~\cite{van2014element,kumar2021topologyDTU}. Further, two members of the mechanisms may also come in contact, i.e., self-contact situations may arise~\cite{kumar2019computational}. These challenges become more pronounced with design-dependent pressure loads, which requires a dedicated in-depth study, which is out of the scope of this paper. In this regard, we assume a small-deformation theory for the presented work.
				
				Hammer and Olhoff~\cite{hammer2000topology} presented the first TO approach for loadbearing structures considering the design-dependent behavior of the pressure loads. One can find the current state-of-the-art for the existing methods wherein most are proposed for designing loadbearing structures in~\cite{picelli2019topology,kumar2020topology,kumar2021topology}. A fictitious thermal model was presented by~\cite{chen2001topology}, which was used in~\cite{chen2001advances} for designing pressure-actuated CMs. Panganiban et al.~\cite{panganiban2010topology} proposed a non-conforming FE-based method for the pressure-actuated CMs. de Souza and Silva~\cite{de2020topology} used the mixed displacement-pressure FEs method with a projection technique; the formulation was also used in~\cite{sigmund2007topology} to design pressure-loaded structures. Kumar et al.~\cite{kumar2020topology} presented an approach using the Darcy law to design pressure-loaded structures and pressure-actuated CMs. The method is extended in~\cite{kumar2021topology} for generating 3D pressure-activated CMs,  with the robust formulation in~\cite{kumar2022topological}, in~\cite{pinskier2023automated} for designing 3D pneumatic multi-material soft grippers, in~\cite{kumar2023TOPress} a 100-line MATLAB code, \texttt{TOPress}, for pressure loadbearing structures and in~\cite{kumar2022improved} with a featured-based TO. The approach uses the standard finite elements and provides consistent load sensitivities using the adjoint-variable method. The moving isosurface threshold (MIST) method is extended for designing pressure-actuated CMs in~\cite{lu2021topology} while neglecting the load sensitivity terms. It is demonstrated in~\cite{kumar2020topology,kumar2023TOPress} that including the load sensitivities alters the final topologies of the optimized designs. Noting the method's efficacy, success, robustness, and generality presented in~\cite{kumar2020topology},  we adopt and modify the approach to model the design-dependent characteristics of the fluidic pressure load herein for multiple materials.

The first method to design compliant mechanisms using TO was presented by Ananthasuresh et al.~\cite{ananthasuresh1994strategies}. They formulated and extremized a weighted objective using the mechanism's flexibility measure (e.g., output displacement) and stiffness measure (e.g., strain energy). Later, a multi-criteria objective was presented in~\cite{frecker1997topological}. Sigmund~\cite{sigmund1997designCM} gave a mechanical advantage maximization-based approach with appropriate input displacement and volume constraints. Typically, the CMs designed using TO suffer from single-node connections~\cite{sigmund1997designCM,yin2003design,saxena2007honeycomb,kumar2022topological}. The robust TO formulation~\cite{wang2011projection} is employed herein to circumvent point connections for the multi-material CMs. Multi-material has seen minimal investigation for pressure-actuated TO, while combining stiff and compliant materials offers exciting potential. We maximize the output deformation of the mechanism as per the robust formulation. The min-max optimization problem is formulated using the output deformation of the eroded and blueprint designs. A constraint on strain energy of the eroded design is considered, whereas volume constraints are applied to the blueprint design~(see Sec.~\ref{Sec:TopOpt_formulation}). 
		
		\PKadd{In summary, the new contributions of the paper are:
		\begin{enumerate}
			\item A TO approach to design pneumatic-actuated multi-material CMs 
			\item Pneumatic load modeling for accounting load's design-dependent characteristics in the multi-material TO setting 
			\item A comparative performance study of optimized CMs  with multiple materials and single material
			\item Optimization of different pneumatically actuated CMs with multiple materials using the proposed method
		\end{enumerate}}

	The remainder of the paper is structured as follows. Sec.~\ref{Sec:Mult-Material modeling} provides the extended SIMP multi-material formulation. Pressure load modeling for the multi-material cases by extending the Darcy law presented in~\cite{kumar2021topology} is described in Sec.~\ref{Sec:PressureLoad modeling}. The consistent nodal loads are also evaluated. Sec.~\ref{Sec:TopOpt_formulation} provides the topology optimization formulation in the robust TO setting. A pseudocode is provided to determine the strain energy's upper limit. Sensitive analysis for the objective and constraints is performed. Numerical results and discussions are provided in Sec.~\ref{Sec:Numerical_Examples_Discussions}. Two- and three-material pneumatically actuated multiple material gripper and contractor mechanisms for different volume fractions are optimized. Lastly, concluding remarks are noted in Sec.~\ref{Sec:Con_remarks}.
				
		
\section{Multi-material modeling}\label{Sec:Mult-Material modeling}
The introduction section introduces some of the existing multi-material models for TO; the readers may refer to Sivapuram et al.~\cite{sivapuram2021design} for a comprehensive list and description. By confining ourselves to density-based TO formulation, we use the extended SIMP approach for multi-material modeling. One writes the modified SIMP for a two-phase (solid and void), i.e., for one material TO as
				    \begin{equation}~\label{Eq:oneMaterial}
				    	E_i = E_v+{\bar{\rho}_i}^p (E_1 - E_v) = (1 -\bar{\rho}_i^p)E_v + \bar{\rho}_i^pE_1,
				    \end{equation} 
 where $E_1$ is Young's modulus of the given material, whereas elastic constant $E_v = 10^{-6}\times E_1$ is assigned to the void elements to avoid the singularity of the structural stiffness matrix during finite element analysis~\cite{sigmund2013topology}. $p$, the SIMP penalty parameter, encourages optimization to 0-1 solutions. $\bar{\rho}_i$ is the physical or projected design variable for element~$i$, which is defined herein as
			    	\begin{equation}\label{Eq:projectionFilt}
			    	\bar{\rho}_i = \frac{\tanh(\beta \eta) + \tanh(\beta(\tilde{\rho}_i-\eta))} {\tanh{\left(\beta \eta\right)}+\tanh{\left(\beta(1 - \eta)\right)}},
			    \end{equation}
		    with  steepness parameter $\beta \in [0,\,\infty)$ and $\eta$ determines the transition point. For all practical purposes, $\beta$ varies from $1$ to a finite number, $\beta_\text{max}$ using a continuation scheme. $\tilde{\rho}_i$, the filtered design variable~~\cite{bourdin2003design}, is obtained from the actual density variable as
		    \begin{equation}\label{Eq:densityfilter}
		    	\tilde{\rho_i} = \frac{\sum_{j=1}^{Nel} v_j \rho_j w(\mvect{x}_j)}{\sum_{j=1}^{Nel} v_j w(\mvect{x}_j)},
		    \end{equation}
	    where \textit{Nel} is the total number of FEs used to parameterize the design domain. The volume of element~$j$ is indicated by $v_j$. $w(\mvect{x}_j)= \max\left(0,\,1-\frac{||\mvect{x}_i -\mvect{x}_j||}{r_\text{fill}}\right)$, is the weight function, wherein $||(.)||$ is a Euclidean  distance between centroids $\mvect{x}_i$ and $\mvect{x}_j$ of elements  $i$ and $j$, respectively. $r_\text{fill}$ indicates the filter radius.
	    
	    One can extend the modified SIMP (Eq.~\ref{Eq:oneMaterial}) scheme for the two-material, i.e., three-phase (material 1, material 2 and void) cases as~\cite{sivapuram2021design}
	    \begin{equation}~\label{Eq:twoMaterial}
	    	E_i = (1 -\bar{\rho}_{i1}^p)E_v + \bar{\rho}_{i1}^p \left(\left(1-\bar{\rho}_{i2}^p \right)E_1 + \bar{\rho}_{i2}^p E_2\right),
	    \end{equation}
    where $E_1$ and $E_2$ are Young's moduli of material 1 and material 2, respectively. $E_v = 10^{-6}\times\min(E_1,\,E_2)$. Note now element~$i$ is assigned two design variables~$\left\{{\rho}_{i1},\,{\rho}_{i2}\right\}$. The corresponding  physical, $\left\{\bar{\rho}_{i1},\,\bar{\rho}_{i2}\right\}$, and filtered variables, $\left\{\bar{\rho}_{i1},\,\bar{\rho}_{i2}\right\}$, can be determined using Eq.~\ref{Eq:projectionFilt} and Eq.~\ref{Eq:densityfilter}, respectively. $\bar{\rho}_{i1} =1$ gives the solid state, whereas $\bar{\rho}_{i1} =0$ indicates the void phase of element~$i$. $\bar{\rho}_{i1} =1$ and $\bar{\rho}_{i2} =0$ imply material 1, whereas $\bar{\rho}_{i1} =1$ and $\bar{\rho}_{i2} =1$ give material 2. Thus, $\bar{\rho}_{i1}$ and $\bar{\rho}_{i2}$ are termed topology and material variables, respectively. Likewise, one writes the extended SIMP scheme for three-material/four-phase cases as
    \begin{equation}~\label{Eq:threeMaterial}
    	E_i = (1 -\bar{\rho}_{i1}^p)E_v + \bar{\rho}_{i1}^p \left(\left(1-\bar{\rho}_{i2}^p \right)E_1 + \bar{\rho}_{i2}^p \left( \left(1-\bar{\rho}_{i3}^p \right)E_2 + \bar{\rho}_{i3}^p E_3\right)\right),
    \end{equation}
where $E_v = 10^{-6}\times\min(E_1,\,E_2,\,E_3)$. $E_i|_{i=1,\,2,\,3}$, are Young's moduli of material~$i$. $\left\{\bar{\rho}_{i1}, \,\bar{\rho}_{i2},\,\bar{\rho}_{i3}  \right\}$ = $\left\{1,\,0,\,0\right\}$, $\left\{\bar{\rho}_{i1}, \,\bar{\rho}_{i2},\,\bar{\rho}_{i3}  \right\}$=$\left\{1,\,1,\,0\right\}$ and $\left\{\bar{\rho}_{i1}, \,\bar{\rho}_{i2},\,\bar{\rho}_{i3}  \right\}$= $\left\{1,\,1,\,1\right\}$ give material~1, material~2 and material~3, respectively. One can readily write the extended SIMP formulation for $m$ candidate materials~\cite{sivapuram2021design}. For $m$ candidate materials, each element in the parameterized domain is assigned $m$ design variables. 


\section{Pressure load modeling}\label{Sec:PressureLoad modeling}
This section extends the pressure load modeling approach presented in~\cite{kumar2020topology} to the multi-material cases. For a detailed overview, the readers can refer to~\cite{kumar2020topology,kumar2021topology}. Ideally, elements achieve solid or void states at the final stage of TO. However, elements can be considered porous in the beginning stage of optimization. Moreover, we have known pressure differences from the given pressure boundary conditions. Therefore, as proposed in~\cite{kumar2020topology}, the Darcy law is used to model the pressure load, wherein the Darcy flux is defined as
\begin{equation}\label{Eq:Darcyflux}
	\bm{q} = -\frac{\kappa}{\mu}\nabla p = -K(\bar{\rho}) \nabla p,
\end{equation}
where $\kappa$, $\mu$, and $\nabla p$ are the permeability of the medium, the fluid viscosity, and the pressure gradient. $K(\bar{\rho})$ is called the flow coefficient, which is defined for element~$i$ in a multi-material setting as
\begin{equation}\label{Eq:Flowcoefficient}
	K(\bar{\rho}_i) = K_v\left(1-(1-\epsilon) \mathcal{H}(\bar{\rho}_{i1},\,\beta_\kappa,\,\eta_\kappa)\right),
\end{equation}
where
$\mathcal{H}(\bar{{\rho}}_{i1},\,\beta_\kappa,\,\eta_\kappa) = \frac{\tanh{\left(\beta_\kappa\eta_\kappa\right)}+\tanh{\left(\beta_\kappa(\bar{\rho}_{i1} - \eta_\kappa)\right)}}{\tanh{\left(\beta_\kappa \eta_\kappa\right)}+\tanh{\left(\beta_\kappa(1 - \eta_\kappa)\right)}}$. $\epsilon = \frac{K_s}{K_v}$, where $K_s$ and $K_v$ are the flow coefficient for the solid and void phases, respectively. $\left\{ \eta_\kappa,\,\beta_\kappa\right\}$  are the flow parameters~\cite{kumar2020topology}. The flow coefficient does not differentiate between different materials, as any candidate material implies a solid phase. That means that the flow coefficient is only related to the topology variable, i.e., $\bar{{\rho}}_{i1}$ as written in~Eq.~\ref{Eq:Flowcoefficient}.

Per~\cite{kumar2020topology,kumar2021topology}, to achieve a realistic pressure drop, a volumetric drainage term, $Q_\text{drain} = -D(\tilde{\rho}_i) (p - p_{\text{ext}})$ is included in the Darcy law (Eq.~\ref{Eq:Darcyflux}). $D(\tilde{\rho}_i)$ is also defined using a Heaviside function~\cite{kumar2020topology}. $p_\text{ext}$, the external pressure, is assumed to be zero. The final balanced equation using the finite element method for the Darcy law with the drainage term is~\cite{kumar2020topology,kumar2021topology}
\begin{equation}\label{Eq:FinalbalanceEqu}
	\mvect{Ap} = \mvect{0},
\end{equation}
where $\mvect{A}$ and $\mvect{p}$ are the global flow matrix and pressure vector respectively. The final pressure field is converted to the consistent nodal forces using
\begin{equation}\label{Eq:nodalforce}
	\mvect{F} = -\mvect{T}\mvect{p},
\end{equation}
where $\mvect{T}$ and $\mvect{F}$ are the global transformation matrix and force vector, respectively. 

\vspace{-0.46cm}
\section{Topology optimization formulation}\label{Sec:TopOpt_formulation}
In this section, we provide the optimization problem formulation and sensitivity analysis. We use the robust formulation~\cite{wang2011projection} with the eroded $\bar{\rho}_e$ and blueprint $\bar{\rho}_b$ variables.  One replaces $\eta$ by $0.5+\Delta\eta$ and $0.5$ in Eq.~\ref{Eq:projectionFilt} for evaluating the eroded and blueprint variables, respectively. $\Delta\eta \in [0,\,0.5]$ is a user-defined parameter. A non-smooth min-max optimization problem is formulated and solved using the method of moving asymptotes~(MMA, cf.~\cite{svanberg1987}). 
We consider the output displacements of the eroded and blueprint (intermediate) designs within the optimization formulation. The volume constraint is applied to the blueprint designs, indicating that the optimized mechanisms are only robust with respect to over-etching~\cite{wang2011projection,kumar2022topological}. 

\subsection{Optimization problem} 
In the robust formulation setting, the optimization problem for the pressure-driven multi-material CMs can be written as:
\begin{equation}\label{Eq:Optimizationequation}
	\begin{rcases}
		& \underset{\bar{\bm{\rho}}(\tilde{\bm{\rho}}(\bm{\rho}))}{\text{min}:}
		&&f_0 = \max_{r}\, {u^\text{out}_r} = \max_{r}\left\{\bm{l}^\top \mathbf{u}_r\right\}|_{r =e,\,b}\\
		& \text{such that:} \,\, &&\,\,\,\,{^1}{\bm{\lambda}_r}:\,\, \mathbf{A}_r\mathbf{p}_r = \mathbf{0 }\\
		&  &&\,{^2}\bm{\lambda}_r:\,\,  \mathbf{K}_r\mathbf{u}_r = \mathbf{F}_r = -\mathbf{T} \mathbf{p}_r\\
		& &&\,{\Lambda}_b^q:\,\,  \text{Volume constraints  on blueprint design} \\
		& && \,{\Lambda}_e:\,\,  \text{g}_2= \frac{SE^e}{SE^*} \le 1\\
		& &&\, 0\le \rho_i\le 1 \,\,\forall i
	\end{rcases},
\end{equation}
where ${r =e,b}$ indicates the eroded and intermediate/blueprint designs, respectively~\cite{wang2011projection,kumar2022topological}. $\mathbf{K}_r$ and $\mathbf{u}_r$ are the global stiffness matrix and global displacement vector, respectively. $u^\text{out}_r$ is the output deformation of the mechanisms in the given direction. $\bm{l}$ is a vector with all zeros except the entry corresponding to the output degree of freedom, which is set to one. $SE^*$ is the desired strain energy of the multi-material mechanisms for the eroded design. $SE^e$ denotes the strain energy of the eroded mechanism. Instead of applying material volume constraints using variable-inseparable expressions~\cite{sigmund1997design}, we use a linear form of that. ${^i}\bm{\lambda}_r|_{i = 1,\,2}$,\, $\Lambda_b^q$ and $\Lambda_e$ are the Lagrange multipliers, where $q$ indicates the number of volume constraints. The volume constraints for the blueprint design for three-phase cases are noted below ($q =2$):
\begin{equation}
\begin{split}
		\displaystyle \sum_{i=1}^{{Nel}}v_i\bar{\rho}_{i1}\le \left(v_{f_1} + v_{f_2}\right) \sum_{i=1}^{{Nel}}v_i,\quad
  \displaystyle \sum_{i=1}^{{Nel}}v_i\bar{\rho}_{i2}\le v_{f_2}\sum_{i=1}^{{Nel}}v_i.
\end{split}
\end{equation}
The first volume constraint limits the material~1 and material~2, whereas the second limits the amount of material~2.
Likewise, for the three-material cases, the volume constraints are ($q =2$):
\begin{equation}
	\begin{split}
		&\displaystyle \sum_{i=1}^{{Nel}}v_i\bar{\rho}_{i1}\le \left(v_{f_1} + v_{f_2} +  v_{f_3}\right) \sum_{i=1}^{{Nel}}v_i,\\
		&\displaystyle \sum_{i=1}^{{Nel}}v_i\bar{\rho}_{i2}\le v_{f_2}\sum_{i=1}^{{Nel}}v_i,\quad
			\displaystyle \sum_{i=1}^{{Nel}}v_i\bar{\rho}_{i3}\le v_{f_3}\sum_{i=1}^{{Nel}}v_i,
	\end{split}
\end{equation}
where $v_{f_1}$, $v_{f_2}$ and $v_{f_3}$ are volume fraction for material~1, material~2 and material~3, respectively. $v_i$ represents volume of element~$i$.

To define $g_2$ (Eq.~\ref{Eq:Optimizationequation}), one needs to provide the upper limit, i.e., $SE^*$. We decide $SE^*$  herein at the initial stage of optimization, i.e., at the first loop of the optimization. The idea is that we should get a realizable load-sustaining design with high performance. The pseudo-code for finding $SE^*$  and defining $g_2$ is given below:

\begin{algorithm}
	\caption{Calculation of $SE^*$ and $g_2$ formulation}
	\begin{algorithmic} 
		\REQUIRE $loop,\, \mathbf{u}_e,\, \mathbf{K}_e,\,SE^e = \frac{1}{2}\mathbf{u}_e^\top\mathbf{K}_e \mathbf{u}_e$
		\IF{$loop==1$}
		\IF{$SE^e-\lfloor SE\rfloor\le 0.5$}                
		\STATE $SE^* = \lfloor SE^e\rfloor$ 
		\ELSE
		\STATE $SE^* = \lfloor SE^e\rfloor + 0.5$ 
		\ENDIF
		\ENDIF
		\STATE $g_2 = \frac{SE^e}{SE^*}\le 1$
	\end{algorithmic}
\end{algorithm}
\vspace{-1cm}
\noindent where $\lfloor SE^e\rfloor$ indicates the greatest integer of $SE_e$.
\subsection{Sensitivity analysis}
The optimization problem in Eq.~\ref{Eq:Optimizationequation} is solved using a gradient-based optimizer, the method of moving asymptotes (MMA)~\cite{svanberg1987}. Thus, the derivatives of the objective and constraints with respect to the design variables are needed, which are obtained herein using the adjoin-variable method herein. One writes the augmented performance $\mathcal{\phi}$ for the objective function  as
\begin{equation}\label{Eq:augmentedperformance}
	\mathcal{\phi} = f_0 + \trr{{^1}\bm{\lambda}}_r ( \mathbf{A}_r\mathbf{p}_r) +  \trr{{^2}\bm{\lambda}}_r \left(\mathbf{K}_r\mathbf{u}_r +\mathbf{T} \mathbf{p}_r\right),
\end{equation}
 Differentiating Eq.~\ref{Eq:augmentedperformance} with respect to the physical design variable~$\bar{\bm{\rho}}_r$,  yields
\begin{equation}\label{Eq:Lagrangeder}
	\begin{split}
		\frac{d \mathcal{\phi}}{d \bar{\bm{\rho}}_r} = &\frac{\partial f_0}{\partial \bar{\bm{\rho}}_r} + 
		\frac{\partial f_0}{\partial\mathbf{u}_r}\frac{\partial\mathbf{u}_r}{\partial \bar{\bm{\rho}}_r}+ {^1}\bm{\lambda}_r^\top \left(\frac{\partial\mathbf{A}_r}{\partial\bar{\bm{\rho}}_r}\mathbf{p}_r\right) + {^1}\bm{\lambda}_r^\top\left(\mathbf{A}_r\frac{\partial\mathbf{p}_r}{\partial\bar{\bm{\rho}}_r}\right)\\ &+ {^2}\bm{\lambda}_r^\top \left(\frac{\partial\mathbf{K}_r}{\partial \bar{\bm{\rho}}_r}\mathbf{u}_r + \mathbf{K}_r\frac{\partial\mathbf{u}_r}{\partial \bar{\bm{\rho}}_r}\right) + {^2}\bm{\lambda}_r^\top \left(\frac{\partial\mathbf{T}}{\partial \bar{\bm{\rho}}_r}\mathbf{p}_r + \mathbf{T}\frac{\partial\mathbf{p}_r}{\partial \bar{\bm{\rho}}_r}\right)\\
		=&  \frac{\partial f_0}{\partial \bar{\bm{\rho}}_r}+{^2}\bm{\lambda}_r^\top \left(\frac{\partial\mathbf{K}_r}{\partial \bar{\bm{\rho}}_r}\mathbf{u}_r\right) + {^1}\bm{\lambda}_r^\top \left(\frac{\partial\mathbf{A}_r}{\partial\bar{\bm{\rho}}_r}\mathbf{p}_r\right) \\ &+ \underbrace{\left(\bm{l}^\top + {^2}\bm{\lambda}_r^\top \mathbf{K}_r\right)}_{\Theta_1}\frac{\partial\mathbf{u}_r}{\partial \bar{\bm{\rho}}_r} + \underbrace{\left({^1}\bm{\lambda}_r^\top\mathbf{A}_r + {^2}\bm{\lambda}_r^\top \mathbf{T} \right)}_{\Theta_2}\frac{\partial\mathbf{p}_r}{\partial\bar{\bm{\rho}}_r}.
	\end{split}
\end{equation} 
Using the fundamentals of the adjoint-variable method, we choose ${^1}\bm{\lambda}_r$ and  ${^2}\bm{\lambda}_r$ such that $\Theta_1 = 0$ and $\Theta_2 = 0$, which give
\begin{equation} \label{Eq:LagrangeMultiplier}
	\begin{aligned}
		{^2}\bm{\lambda}_r^\top = -\bm{l}^\top\inv{\mathbf{K}}_r,\quad
		{^1}\bm{\lambda}_r^\top = -{^2}\bm{\lambda}_r^\top \mathbf{T} \inv{\mathbf{A}}_r = \bm{l}^\top\inv{\mathbf{K}}_r\mathbf{T} \inv{\mathbf{A}}_r,
	\end{aligned}
\end{equation}
\begin{figure*}[h!]
	\centering
	\begin{subfigure}[t]{0.450\textwidth}
		\centering
		\includegraphics[scale=0.9]{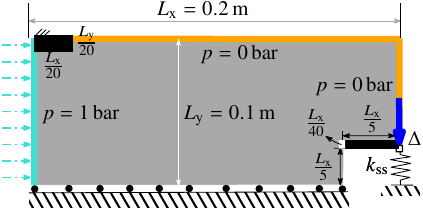}
		\caption{}
		\label{fig:GPdesign}
	\end{subfigure}
	\begin{subfigure}[t]{0.450\textwidth}
		\centering
		\includegraphics[scale=0.9]{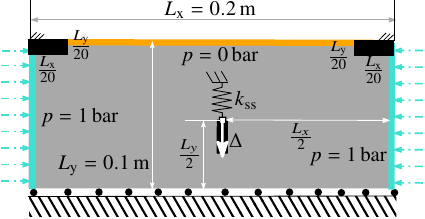}
		\caption{}
		\label{fig:ConDesign}
	\end{subfigure}
	\caption{Symmetric half-design domains for the mechanisms. (\subref{fig:GPdesign}) Gripper design domain and (\subref{fig:ConDesign}) Contractor design domain. $L_x = \SI{0.2}{\meter}$ and $L_y =\SI{0.1}{\meter}$, where $L_x$ and $L_y$ are the dimensions in $x-$ and $y-$directions, respectively. The pneumatic input load is applied from the left edge of the gripper domain, whereas it is applied from the left and right edges for the contractor mechanisms. The workpiece is indicated by spring with stiffness $k_\text{ss}$. Solid and void non-design domains are also depicted. $\Delta$ is the output deformation.} \label{fig:GP_Con_Design}
\end{figure*}

\begin{figure*}[h!]
	\centering
	\begin{subfigure}[t]{0.3\textwidth}
		\centering
		\includegraphics[scale=0.5]{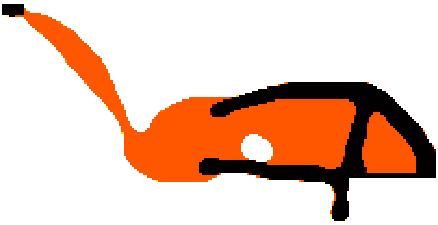}
		\caption{$u_\text{out}= -\SI{5.9}{\milli\meter}$}
		\label{fig:GP_MM2_half}
	\end{subfigure}
	\begin{subfigure}[t]{0.3\textwidth}
		\centering
		\includegraphics[scale=0.45]{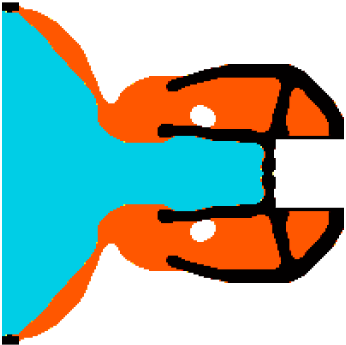}
		\caption{}
		\label{fig:GP_MM2_full_undeform}
	\end{subfigure}
	\begin{subfigure}[t]{0.3\textwidth}
	\centering
	\includegraphics[scale=0.45]{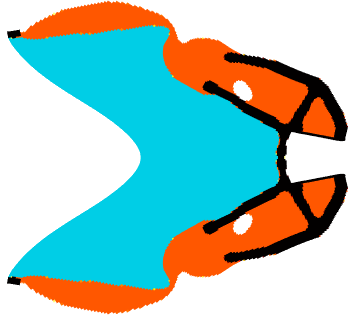}
	\caption{}
	\label{fig:GP_MM2_full_deform}
\end{subfigure}
	\caption{Optimized gripper mechanisms with  two materials (\subref{fig:GP_MM2_half}) Symmetric half optimized gripper mechanism (\subref{fig:GP_MM2_full_undeform}) Full optimized gripper mechanism with final pressure field and (\subref{fig:GP_MM2_full_deform}) Deformed profile of full optimized gripper mechanism with final pressure field. Red$\to$ Material~1 (Young's modulus =$\SI{1e7}{\newton\per\square\meter})$, Black$\to$Material~2 (Young's modulus = $\SI{1e8}{\newton\per\square\meter})$. The output displacement {$u_\text{out}$ is in $y$ direction.}} \label{fig:GP_MM2_all}
\end{figure*}

\begin{figure*}[h!]
	\centering
	\begin{subfigure}[t]{0.3\textwidth}
		\centering
		\includegraphics[scale=0.5]{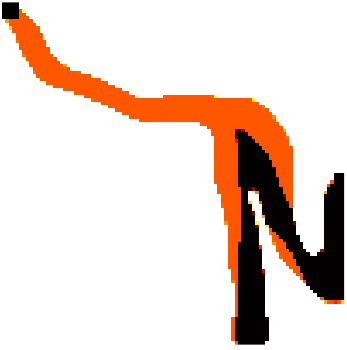}
		\caption{$u_\text{out}= -\SI{2.8}{\milli\meter}$}
		\label{fig:CoN_MM2_half}
	\end{subfigure}
	\begin{subfigure}[t]{0.3\textwidth}
		\centering
		\includegraphics[scale=0.45]{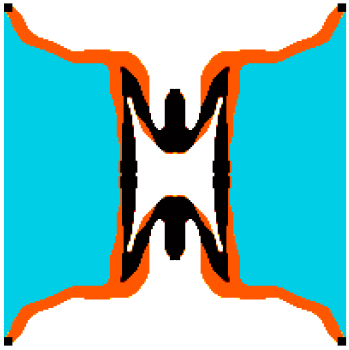}
		\caption{}
		\label{fig:CoN_MM2_full_undeform}
	\end{subfigure}
	\begin{subfigure}[t]{0.3\textwidth}
		\centering
		\includegraphics[scale=0.45]{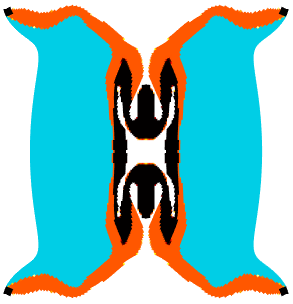}
		\caption{}
		\label{fig:CoN_MM2_full_deform}
	\end{subfigure}
	\caption{Optimized contractor mechanisms with  two materials (\subref{fig:CoN_MM2_half}) Symmetric half optimized contractor mechanism  (\subref{fig:CoN_MM2_full_undeform}) Full optimized contractor mechanism with final pressure field and (\subref{fig:CoN_MM2_full_deform}) Deformed profile of full optimized contractor mechanism with final pressure field. Red$\to$ Material~1 (Young's modulus =$\SI{1e7}{\newton\per\square\meter})$, Black$\to$Material~2 (Young's modulus = $\SI{1e8}{\newton\per\square\meter})$. The output displacement {$u_\text{out}$ is in $y$ direction.}} \label{fig:CoN_MM2_all}
\end{figure*}

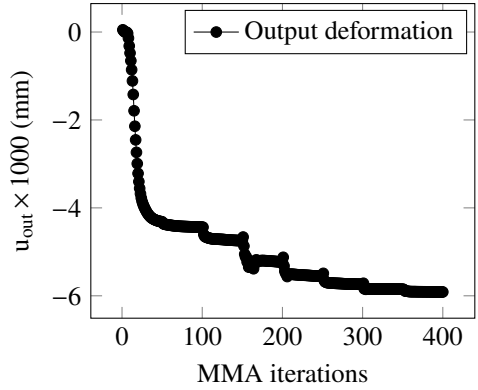
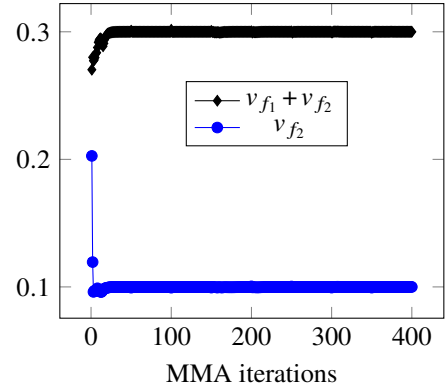
\begin{figure*}
	\centering
	\begin{subfigure}[t]{0.45\textwidth}
		\centering
		\begin{tikzpicture} 	
			\pgfplotsset{compat = 1.3}
			\begin{axis}[
				width = 0.78\textwidth,
				xlabel=MMA iterations,
				ylabel= u$_\text{out}\times 1000$ ($\si{\milli \meter}$)]
				\pgfplotstableread{GPMM2obj.txt}\mydata;
				\addplot[mark=*,black]
				table {\mydata};
				\addlegendentry{Output deformation}
			\end{axis}
		\end{tikzpicture}
		\caption{Objective convergence history}
		\label{fig:gripperobjhistory}
	\end{subfigure}
	\begin{subfigure}[t]{0.45\textwidth}
		\centering
		\begin{tikzpicture} 	
			\pgfplotsset{compat = 1.3}
			\begin{axis}[
				width = 0.78\textwidth,
				xlabel=MMA iterations,
				ylabel=  ,
				legend style={at={(0.75,0.65)},anchor=east}]
				\pgfplotstableread{GPMM2vf1.txt}\mydata;
				\addplot[mark=diamond*,black]
				table {\mydata};
				\addlegendentry{$v_{f_1}+v_{f_2}$}
				\pgfplotstableread{GPMM2vf2.txt}\mydata;
				\addplot[mark=otimes*,blue]
				table {\mydata};
				\addlegendentry{$v_{f_2}$ }
			\end{axis}
		\end{tikzpicture}
		\caption{Convergence plots of volume constraints}
		\label{fig:grippervolumehistory}
	\end{subfigure}
	\caption{Convergence plots of the objective and volume constraints for the two-material gripper mechanism.}\label{fig:gripperconvergence}
\end{figure*}

\begin{figure*}[h!]
	\centering
	\begin{subfigure}[t]{0.3\textwidth}
		\centering
		\includegraphics[scale=0.5]{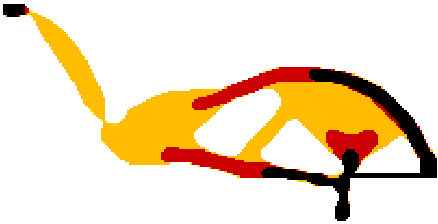}
		\caption{$u_\text{out}= -\SI{8.2}{\milli\meter}$}
		\label{fig:GP_MM3_half}
	\end{subfigure}
	\begin{subfigure}[t]{0.3\textwidth}
		\centering
		\includegraphics[scale=0.45]{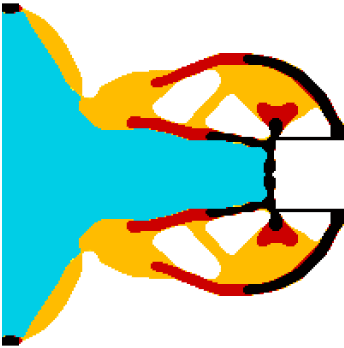}
		\caption{}
		\label{fig:GP_MM3_full_undeform}
	\end{subfigure}
	\begin{subfigure}[t]{0.3\textwidth}
		\centering
		\includegraphics[scale=0.45]{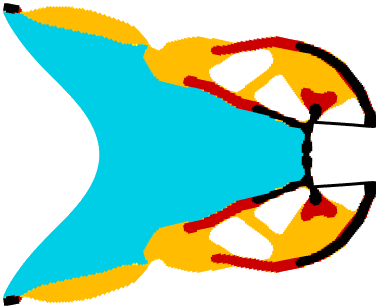}
		\caption{}
		\label{fig:GP_MM3_full_deform}
	\end{subfigure}
	\caption{Optimized gripper mechanisms with  three materials (\subref{fig:GP_MM3_half}) Symmetric half optimized gripper mechanism (\subref{fig:GP_MM3_full_undeform}) Full optimized gripper mechanism with final pressure field and (\subref{fig:GP_MM3_full_deform}) Deformed profile of full optimized gripper mechanism with final pressure field.Yellow$\to$ Material~1 (Young's modulus =$\SI{1e7}{\newton\per\square\meter})$, red$\to$Material~2 (Young's modulus =$0.5\SI{1e8}{\newton\per\square\meter})$ and black$\to$Material~3 (Young's modulus =$\SI{1e8}{\newton\per\square\meter})$. The output displacement {$u_\text{out}$ is in $y$ direction.}} \label{fig:GP_MM3_all}
\end{figure*}

\begin{figure*}[h!]
	\centering
	\begin{subfigure}[t]{0.3\textwidth}
		\centering
		\includegraphics[scale=0.5]{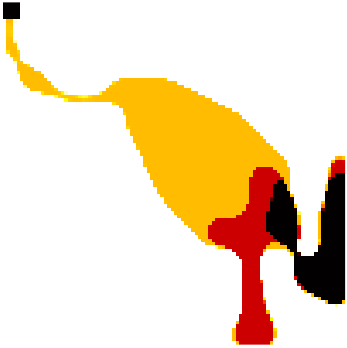}
		\caption{$u_\text{out}= -\SI{4.0}{\milli\meter}$}
		\label{fig:CoN_MM3_half}
	\end{subfigure}
	\begin{subfigure}[t]{0.3\textwidth}
		\centering
		\includegraphics[scale=0.45]{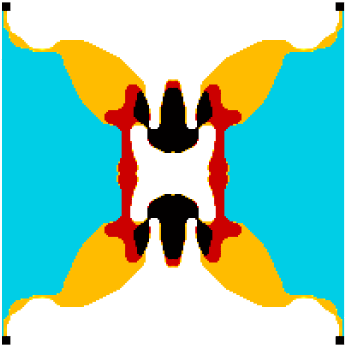}
		\caption{}
		\label{fig:CoN_MM3_full_undeform}
	\end{subfigure}
	\begin{subfigure}[t]{0.3\textwidth}
		\centering
		\includegraphics[scale=0.45]{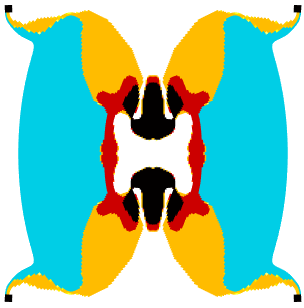}
		\caption{}
		\label{fig:CoN_MM3_full_deform}
	\end{subfigure}
	\caption{Optimized contractor mechanisms with  three materials (\subref{fig:CoN_MM2_half}) Symmetric half optimized contractor mechanism  (\subref{fig:CoN_MM2_full_undeform}) Full optimized contractor mechanism with final pressure field and (\subref{fig:CoN_MM2_full_deform}) Deformed profile of full optimized contractor mechanism with final pressure field. Red$\to$ Material~1 (Young's modulus =$\SI{1e7}{\newton\per\square\meter})$, Black$\to$Material~2 (Young's modulus = $\SI{1e8}{\newton\per\square\meter})$. The output displacement {$u_\text{out}$ is in $y$ direction.}} \label{fig:CoN_MM3_all}
\end{figure*}

and, noting that, $\frac{\partial f_0}{\partial \bar{\bm{\rho}}_r}=0$, one gets
\begin{equation}\label{Eq:senstivities_Obj}
	\frac{d {f_0}}{d \bar{\bm{\rho}}_r} = -\bm{l}^\top\inv{\mathbf{K}}_r \frac{\partial\mathbf{K}_r}{\partial \tilde{\bm{\rho}}_r}\mathbf{u}_r  \underbrace{+\bm{l}^\top\inv{\mathbf{K}}_r\mathbf{T} \inv{\mathbf{A}}_r\frac{\partial\mathbf{A}_r}{\partial\bar{\bm{\rho}}_r}\mathbf{p}_r}_{\text{Load sensitivities}}.
\end{equation}
\begin{figure*}[h!]
	\centering
	\quad
	\begin{subfigure}[t]{0.30\textwidth}
		\centering
		\includegraphics[scale=0.45]{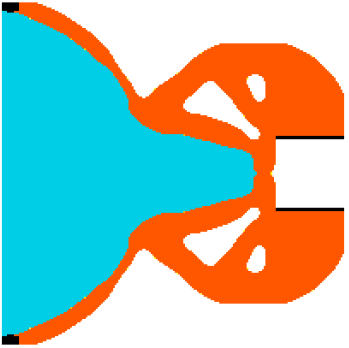}
		\caption{$u_\text{out}= -\SI{3.4}{\milli\meter}$}
		\label{fig:GP_M1_full}
	\end{subfigure}
	\begin{subfigure}[t]{0.30\textwidth}
		\centering
		\includegraphics[scale=0.45]{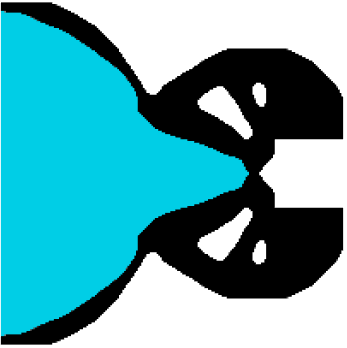}
		\caption{$u_\text{out}= -\SI{2.4}{\milli\meter}$}
		\label{fig:GP_M2_full}
	\end{subfigure}
	\begin{subfigure}[t]{0.30\textwidth}
		\centering
		\includegraphics[scale=0.45]{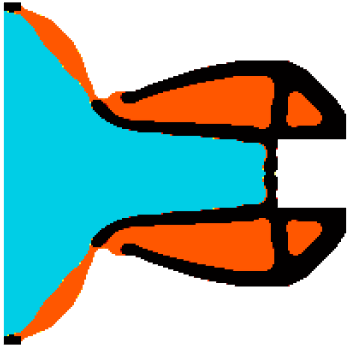}
		\caption{$u_\text{out}= -\SI{6.5}{\milli\meter}$}
		\label{fig:GP_M1M2_full}
	\end{subfigure}
	\caption{ Optimized results for the gripper mechanisms for different cases. (\subref{fig:GP_M1_full}) CASE~I, (\subref{fig:GP_M2_full}) CASE~II, (\subref{fig:GP_M1M2_full}) CASE~III. Red$\to$ Material~1 (Young's modulus =$\SI{1e7}{\newton\per\square\meter})$, black$\to$Material~2 (Young's modulus =$\SI{1e8}{\newton\per\square\meter})$}  \label{fig:GP_M1M2}
\end{figure*} 

Likewise, following the above steps, one finds the derivatives of $\text{g}_2$ (Eq.~\ref{Eq:Optimizationequation}) with respect to the physical variables as
\begin{equation}\label{Eq:senstivities_Obj1}
	\frac{d {\text{g}_2}}{d \bar{\bm{\rho}}_r} = \frac{-\frac{1}{2}\mathbf{u}^\top_r \frac{\partial\mathbf{K}_r}{\partial \bar{\bm{\rho}}}\mathbf{u}_r + \mathbf{u}^\top_r \mathbf{T} \mathbf{A}^{-1}_r\frac{\partial\mathbf{A}_r}{\partial\bar{\bm{\rho}}_r}\mathbf{p}_r}{SE^*},
\end{equation}
Now, one uses the chain rule to determine the derivatives of the objective and constraints with respect to the design variable. Say, $f$ is a representative function for the objective and constraints; then the chain rule can be written as
\begin{equation}\label{Eq:objderivative}
	\frac{d {f}}{d \bm{\rho}}_r = \frac{d {f}}{d \bar{\bm{\rho}}_r}\frac{d \bar{\bm{\rho}}_r}{d \tilde{\bm{\rho}}_r} \frac{d \tilde{\bm{\rho}}_r}{d \bm{\rho}_r}.
\end{equation}
With all ingredients discussed above, we next present numerical examples and discussions.

\section{Numerical examples and discussions}\label{Sec:Numerical_Examples_Discussions}
This section presents pressure-actuated multi-material gripper and contractor  CMs to demonstrate the efficacy of the proposed design approach and to explore the large design space offered by multi-material topology optimization for this problem. The symmetric half-design domains are depicted in Fig.~\ref{fig:GP_Con_Design}. $L_x = \SI{0.2}{\meter}$ and $L_y=\SI{0.1}{\meter}$ are taken, where $L_x$ and $L_y$ indicate the dimension of the mechanisms in the $x-$ and $y-$directions, respectively. The displacement and symmetry boundary conditions are displayed in the figure. The gripper mechanism is actuated from the left side, whereas the contractor mechanism is activated from both the left and right sides. To facilitate gripping of a workpiece with spring constant $k_{ss} =\SI{5e4}{\newton\per\meter}$, a void area of size $\frac{L_x}{5} \times \frac{L_y}{5}$ and a solid region of area $\frac{L_x}{5} \times \frac{L_y}{50}$ are considered. At the fixed boundary conditions, non-design solid regions with dimension $\frac{L_x}{20} \times \frac{L_y}{20}$ are considered (Fig.~\ref{fig:GP_Con_Design}).

The applied input (excess) pressure load is set to $\SI{1}{\bar}$ (Fig.~\ref{fig:GP_Con_Design}). $\SI{0}{\bar}$ pressure is set to the remaining edges excluding the symmetric ones. $N_\text{ex}\times N_\text{ey} = 200\times 100$ elements are used to parameterize the gripper design, whereas  $N_\text{ex}\times N_\text{ey} = 100\times 100$ elements are taken to describe the one-quarter of the contractor design domain. $N_\text{ex}$ and $N_\text{ey}$ represent the number of the bi-linear quadrilateral elements in $x-$ and $y-$directions, respectively. Filter radius, $r_\text{min} = 8.4\times \min(\frac{L_x}{N_\text{ex}}, \frac{L_y}{N_\text{ey}})$ is set. Poisson's ratio $\nu = 0.4$ is used with plane-stress assumptions. The out-of-plane thickness is set to $\SI{0.01}{\meter}$. $\left\{\beta_k,\,\eta_k\right\}$ = $\left\{\beta_k,\,\eta_k\right\}$ = $\left\{10,\,0.1\right\}$ is taken. The external move limit for the MMA optimizer is set to $0.1$. The maximum number of MMA iterations is set to 400. For the projection filter (Eq.~\ref{Eq:projectionFilt}), $\beta$ is updated from 1 to 128 in a continuation manner wherein $\beta$ is doubled after every 50 MMA iterations, and once it gets to 128, it remains so for the remaining iterations.

\subsection{Two-material mechanisms}\label{Sec:Two-material}
Herein two candidate materials with $E_1 =\SI{1e7}{\newton\per\square\meter}$ and $E_2=\SI{1e8}{\newton\per\square\meter}$ are considered for designing pneumatic-activated gripper and contractor mechanisms. $E_v = \SI{1e-6}{}\times\min(E_1,\,E_2)$. $v_{f_{1}} = 0.2$ and $v_{f_{2}} = 0.1$ are considered for the gripper mechanism, whereas $v_{f_{1}} = 0.1$ and $v_{f_{2}} = 0.1$ are set for the contractor mechanism. $\Delta\eta$~\cite{kumar2022topological,wang2011projection} for the gripper and contractor mechanisms are set to 0.05 and 0.15, respectively. 

The symmetric half results are shown in Fig.~\ref{fig:GP_MM2_half} and Fig.~\ref{fig:CoN_MM2_half} for the gripper and contractor mechanisms, respectively. Their complete parts (undeformed) are obtained by suitably transforming the symmetric half-optimized mechanisms and are depicted in Fig.~\ref{fig:GP_MM2_full_undeform} and Fig.~\ref{fig:CoN_MM2_full_undeform}, respectively. The working of these mechanisms is depicted by using their deformed profiles in Fig.~\ref{fig:GP_MM2_full_deform} and Fig.~\ref{fig:CoN_MM2_full_deform}, respectively. The output deformation for the gripper and contractor mechanisms in the desired directions are noted $\SI{5.9}{\milli\meter}$ and $\SI{2.8}{\milli\meter}$, respectively. The objective and volume constraints convergence curves for the gripper mechanism are depicted in Fig.~\ref{fig:gripperconvergence}. The steps in the objective convergence plot (Fig.~\ref{fig:gripperobjhistory}) are due to $\beta$ updation. The volume constraints are satisfied and remain active at the end of the optimization. Both materials constitute the pressure loading boundary. 

\subsection{Three-material mechanisms}
In this section, the mechanisms are designed using three candidate designs.  $E_1 =\SI{1e7}{\newton\per\square\meter}$, $E_2=\SI{0.5e8}{\newton\per\square\meter}$ and $E_3 =\SI{1e8}{\newton\per\square\meter}$ are chosen. $E_v = \SI{1e-6}{}\times\min(E_1,\,E_2,\,E_3)$. $v_{f_{1}} = 0.1$, $v_{f_{2}} = 0.1$ and $v_{f_{1}} = 0.05$ are considered for mechanisms.  $\Delta\eta$~\cite{kumar2022topological,wang2011projection} for the mechanisms are set to 0.01.

The optimized symmetric half results are displayed in Fig.~\ref{fig:GP_MM3_half} and Fig.~\ref{fig:CoN_MM3_half} for the gripper and contractor mechanisms, respectively. The fully optimized results with the final pressure field for the gripper and contractor mechanisms are shown in Fig.~\ref{fig:GP_MM3_full_undeform} and Fig.~\ref{fig:CoN_MM3_full_undeform}, respectively. Fig.~\ref{fig:GP_MM3_full_deform} and Fig.~\ref{fig:CoN_MM3_full_deform} demonstrate the deformed profiles of the gripper and contractor mechanisms, respectively. One can note that the members of the mechanisms can come in contact, i.e., self-contact situation~\cite{kumar2019computational,kumar2022topological}, if  higher pressure loads are applied. The obtained output deformations in the desired directions for the gripper and contractor mechanisms are $\SI{8.2}{\milli\meter} $ and $\SI{4.0}{\milli\meter}$, respectively. 
  
\subsection{Performance comparison}
To gain insight into the question of whether combining different materials offers benefits, this section provides a performance comparison study for the gripper mechanism when it is optimized using one and two materials. The design, optimization, and other parameters are the same as in Sec.~\ref{Sec:Two-material}. We consider three cases: CASE~I: the gripper is designed using material~1 ($E_1 = \SI{1e7}{\newton\per\square\meter}$) and  volume fraction~0.30, CASE~II: the gripper is designed using material~2 ($E_2 = \SI{1e8}{\newton\per\square\meter}$) and volume fraction~0.30 and CASE~III: the gripper is designed using material~1  and material~2 with volume fraction 0.15 for each.

The optimized designs for CASE~I, CASE~II and CASE~III with the final pressure field are depicted in Fig.~\ref{fig:GP_M1_full}, Fig.~\ref{fig:GP_M2_full} and Fig.~\ref{fig:GP_M1M2_full}, respectively. The optimized topology of CASE~I and CASE~II are the same, which should be the case as we consider a linear mechanics setting. The obtained output displacements for CASE~I, CASE~II and CASE~III are $-\SI{3.4}{\milli \meter}$, $-\SI{2.4}{\milli \meter}$ and $-\SI{6.5}{\milli \meter}$. It can be noted that the mechanism of CASE~III, i.e., with two materials, performs clearly better than that of CASE~I and CASE~II. \PKadd{We expect a similar trend with three or more than three materials.} Therefore, the numerical experiment performed here establishes that the use of multiple materials significantly increases the potential performance that can be achieved. In fact, the performance has nearly doubled. It should also be noted that an equal volume constraint was used on both materials, which is an arbitrary choice and which is likely not the optimal ratio. Hence, the true potential of a multi-material design might be even higher.

\section{Concluding remarks}\label{Sec:Con_remarks}
This paper presents a density-based topology optimization approach to design pneumatically actuated multi-material compliant mechanisms. The success of the proposed method is demonstrated by creating two- and three-material gripper and contractor mechanisms. The extended SIMP scheme is employed for modeling multiple materials. Output deformation in the robust TO setting with constraints on volume and strain energy is maximized. The upper limit on the strain energy is evaluated at the beginning of the optimization. The given volume constraints on the material are satisfied and remain active at the end of the optimization.

The Darcy law proposed in~\cite{kumar2021topology} is extended for modeling pressure loads with many candidate materials in a topology optimization setting. The flow coefficient of an element is determined using its solid and void phases' flow coefficient, irrespective of the candidate material type. A standard finite element formulation converts the distributed pressure field into consistent nodal loads. The load sensitivities arising due to the design-dependent nature of the loads are readily determined and considered within the formulation.

The output deformations of the mechanisms are obtained in the desired directions. We demonstrate via numerical experiments that a mechanism optimized using multiple materials performs significantly better than when optimized for a single individual material. Small-deformation finite element analysis has been applied, i.e., nonlinearities due to geometry and material behavior must be addressed. Considering these nonlinearities for more realistic cases \PKadd{with experimental verification} forms the future direction for optimizing such mechanisms.


\begin{thebibliography}{10}

\bibitem{howell2001compliant}
L.~L. Howell, {\em Compliant Mechanisms}.
\newblock John Wiley \& Sons, New York, 2001.

\bibitem{zhu2020design}
B.~Zhu, X.~Zhang, H.~Zhang, J.~Liang, H.~Zang, H.~Li, and R.~Wang, ``Design of
  compliant mechanisms using continuum topology optimization: {A} review,''
  {\em Mechanism and Machine Theory}, vol.~143, p.~103622, 2020.

\bibitem{xavier2022soft}
M.~S. Xavier, C.~D. Tawk, A.~Zolfagharian, J.~Pinskier, D.~Howard, T.~Young,
  J.~Lai, S.~M. Harrison, Y.~K. Yong, and M.~Bodaghi, ``Soft pneumatic
  actuators: A review of design, fabrication, modeling, sensing, control and
  applications,'' {\em IEEE Access}, 2022.

\bibitem{kumar2023towards}
P.~Kumar, ``Towards {T}opology {O}ptimization of {P}ressure-{D}riven {S}oft
  {R}obots,'' in {\em Conference on Microactuators and Micromechanisms},
  pp.~19--30, Springer, 2022.

\bibitem{bandyopadhyay2018additive}
A.~Bandyopadhyay and B.~Heer, ``Additive manufacturing of multi-material
  structures,'' {\em Materials Science and Engineering: R: Reports}, vol.~129,
  pp.~1--16, 2018.

\bibitem{langelaar2016topology}
M.~Langelaar, ``Topology optimization of {3D} self-supporting structures for
  additive manufacturing,'' {\em Additive manufacturing}, vol.~12, pp.~60--70,
  2016.

\bibitem{sigmund2013topology}
O.~Sigmund and K.~Maute, ``Topology optimization approaches,'' {\em Structural
  and Multidisciplinary Optimization}, vol.~48, no.~6, pp.~1031--1055, 2013.

\bibitem{van2013level}
N.~P. Van~Dijk, K.~Maute, M.~Langelaar, and F.~Van~Keulen, ``Level-set methods
  for structural topology optimization: a review,'' {\em Structural and
  Multidisciplinary Optimization}, vol.~48, pp.~437--472, 2013.

\bibitem{thomsen1992topology}
J.~Thomsen, ``Topology optimization of structures composed of one or two
  materials,'' {\em Structural optimization}, vol.~5, pp.~108--115, 1992.

\bibitem{sigmund1997design}
O.~Sigmund and S.~Torquato, ``Design of materials with extreme thermal
  expansion using a three-phase topology optimization method,'' {\em Journal of
  the Mechanics and Physics of Solids}, vol.~45, no.~6, pp.~1037--1067, 1997.

\bibitem{gao2011mass}
T.~Gao and W.~Zhang, ``A mass constraint formulation for structural topology
  optimization with multiphase materials,'' {\em International Journal for
  Numerical Methods in Engineering}, vol.~88, no.~8, pp.~774--796, 2011.

\bibitem{fujii2001composite}
D.~Fujii, B.~Chen, and N.~Kikuchi, ``Composite material design of
  two-dimensional structures using the homogenization design method,'' {\em
  International Journal for Numerical Methods in Engineering}, vol.~50, no.~9,
  pp.~2031--2051, 2001.

\bibitem{yin2001topology}
L.~Yin and G.~K. Ananthasuresh, ``Topology optimization of compliant mechanisms
  with multiple materials using a peak function material interpolation
  scheme,'' {\em Structural and Multidisciplinary Optimization}, vol.~23,
  no.~1, pp.~49--62, 2001.

\bibitem{chu2018stress}
S.~Chu, L.~Gao, M.~Xiao, Z.~Luo, and H.~Li, ``Stress-based multi-material
  topology optimization of compliant mechanisms,'' {\em International Journal
  for Numerical Methods in Engineering}, vol.~113, no.~7, pp.~1021--1044, 2018.

\bibitem{gaynor2014multiple}
A.~T. Gaynor, N.~A. Meisel, C.~B. Williams, and J.~K. Guest,
  ``Multiple-material topology optimization of compliant mechanisms created via
  polyjet three-dimensional printing,'' {\em Journal of Manufacturing Science
  and Engineering}, vol.~136, no.~6, 2014.

\bibitem{zuo2017multi}
W.~Zuo and K.~Saitou, ``Multi-material topology optimization using ordered
  {SIMP} interpolation,'' {\em Structural and Multidisciplinary Optimization},
  vol.~55, no.~2, pp.~477--491, 2017.

\bibitem{sivapuram2021design}
R.~Sivapuram, R.~Picelli, G.~H. Yoon, and B.~Yi, ``On the design of
  multimaterial structural topologies using integer programming,'' {\em
  Computer Methods in Applied Mechanics and Engineering}, vol.~384, p.~114000,
  2021.

\bibitem{hammer2000topology}
V.~B. Hammer and N.~Olhoff, ``Topology optimization of continuum structures
  subjected to pressure loading,'' {\em Structural and Multidisciplinary
  Optimization}, vol.~19, no.~2, pp.~85--92, 2000.

\bibitem{kumar2020topology}
P.~Kumar, J.~S. Frouws, and M.~Langelaar, ``Topology optimization of fluidic
  pressure-loaded structures and compliant mechanisms using the {Darcy}
  method,'' {\em Structural and Multidisciplinary Optimization}, vol.~61,
  no.~4, pp.~1637--1655, 2020.

\bibitem{sigmund2007topology}
O.~Sigmund and P.~M. Clausen, ``Topology optimization using a mixed
  formulation: an alternative way to solve pressure load problems,'' {\em
  Computer Methods in Applied Mechanics and Engineering}, vol.~196, no.~13-16,
  pp.~1874--1889, 2007.

\bibitem{picelli2019topology}
R.~Picelli, A.~Neofytou, and H.~A. Kim, ``Topology optimization for
  design-dependent hydrostatic pressure loading via the level-set method,''
  {\em Structural and Multidisciplinary Optimization}, vol.~60, no.~4,
  pp.~1313--1326, 2019.

\bibitem{kumar2023TOPress}
P.~Kumar, ``{TOPress}: a {MATLAB} implementation for topology optimization of
  structures subjected to design-dependent pressure loads,'' {\em Structural
  and Multidisciplinary Optimization}, vol.~66, no.~4, 2023.

\bibitem{van2014element}
N.~P. van Dijk, M.~Langelaar, and F.~Van~Keulen, ``Element deformation scaling
  for robust geometrically nonlinear analyses in topology optimization,'' {\em
  Structural and Multidisciplinary Optimization}, vol.~50, pp.~537--560, 2014.

\bibitem{kumar2021topologyDTU}
P.~Kumar, C.~Schmidleithner, N.~Larsen, and O.~Sigmund, ``Topology optimization
  and {3D} printing of large deformation compliant mechanisms for straining
  biological tissues,'' {\em Structural and Multidisciplinary Optimization},
  vol.~63, pp.~1351--1366, 2021.

\bibitem{kumar2019computational}
P.~Kumar, A.~Saxena, and R.~A. Sauer, ``Computational synthesis of large
  deformation compliant mechanisms undergoing self and mutual contact,'' {\em
  Journal of Mechanical Design}, vol.~141, no.~1, 2019.

\bibitem{kumar2021topology}
P.~Kumar and M.~Langelaar, ``On topology optimization of design-dependent
  pressure-loaded three-dimensional structures and compliant mechanisms,'' {\em
  International Journal for Numerical Methods in Engineering}, vol.~122, no.~9,
  pp.~2205--2220, 2021.

\bibitem{chen2001topology}
B.~C. Chen and N.~Kikuchi, ``Topology optimization with design-dependent
  loads,'' {\em Finite elements in analysis and design}, vol.~37, no.~1,
  pp.~57--70, 2001.

\bibitem{chen2001advances}
B.~C. Chen, E.~C. Silva, and N.~Kikuchi, ``Advances in computational design and
  optimization with application to {MEMS},'' {\em International Journal for
  Numerical Methods in Engineering}, vol.~52, no.~1-2, pp.~23--62, 2001.

\bibitem{panganiban2010topology}
H.~Panganiban, G.-W. Jang, and T.-J. Chung, ``Topology optimization of
  pressure-actuated compliant mechanisms,'' {\em Finite Elements in Analysis
  and Design}, vol.~46, no.~3, pp.~238--246, 2010.

\bibitem{de2020topology}
E.~M. de~Souza and E.~C.~N. Silva, ``Topology optimization applied to the
  design of actuators driven by pressure loads,'' {\em Structural and
  Multidisciplinary Optimization}, vol.~61, no.~5, pp.~1763--1786, 2020.

\bibitem{kumar2022topological}
P.~Kumar and M.~Langelaar, ``Topological synthesis of fluidic pressure-actuated
  robust compliant mechanisms,'' {\em Mechanism and Machine Theory}, vol.~174,
  p.~104871, 2022.

\bibitem{pinskier2023automated}
J.~Pinskier, P.~Kumar, M.~Langelaar, and D.~Howard, ``Automated design of
  pneumatic soft grippers through design-dependent multi-material topology
  optimization,'' in {\em 6th IEEE-RAS International Conference on Soft
  Robotics (RoboSoft 2023)}, IEEE, 2023.

\bibitem{kumar2022improved}
P.~Kumar and A.~Saxena, ``An improved material{ Mask Overlay Strategy} for the
  desired discreteness of pressure-loaded optimized topologies,'' {\em
  Structural and Multidisciplinary Optimization}, vol.~65, no.~10, p.~304,
  2022.

\bibitem{lu2021topology}
Y.~Lu and L.~Tong, ``Topology optimization of compliant mechanisms and
  structures subjected to design-dependent pressure loadings,'' {\em Structural
  and Multidisciplinary Optimization}, vol.~63, no.~4, pp.~1889--1906, 2021.

\bibitem{ananthasuresh1994strategies}
G.~K. Ananthasuresh, S.~Kota, and N.~Kikuchi, ``Strategies for systematic
  synthesis of compliant mems,'' 1994.

\bibitem{frecker1997topological}
M.~Frecker, G.~K. Ananthasuresh, S.~Nishiwaki, N.~Kikuchi, and S.~Kota,
  ``Topological synthesis of compliant mechanisms using multi-criteria
  optimization,'' 1997.

\bibitem{sigmund1997designCM}
O.~Sigmund, ``On the design of compliant mechanisms using topology
  optimization,'' {\em Journal of Structural Mechanics}, vol.~25, no.~4,
  pp.~493--524, 1997.

\bibitem{yin2003design}
L.~Yin and G.~Ananthasuresh, ``Design of distributed compliant mechanisms,''
  {\em Mechanics based design of structures and machines}, vol.~31, no.~2,
  pp.~151--179, 2003.

\bibitem{saxena2007honeycomb}
R.~Saxena and A.~Saxena, ``On honeycomb representation and {SIGMOID} material
  assignment in optimal topology synthesis of compliant mechanisms,'' {\em
  Finite Elements in Analysis and Design}, vol.~43, no.~14, pp.~1082--1098,
  2007.

\bibitem{wang2011projection}
F.~Wang, B.~S. Lazarov, and O.~Sigmund, ``On projection methods, convergence
  and robust formulations in topology optimization,'' {\em Structural and
  multidisciplinary optimization}, vol.~43, pp.~767--784, 2011.

\bibitem{bourdin2003design}
B.~Bourdin and A.~Chambolle, ``Design-dependent loads in topology
  optimization,'' {\em ESAIM: Control, Optimisation and Calculus of
  Variations}, vol.~9, pp.~19--48, 2003.

\bibitem{svanberg1987}
K.~Svanberg, ``The method of moving asymptotes{--}a new method for structural
  optimization,'' {\em Int J Numer Meth Eng}, vol.~24, no.~2, pp.~359--373,
  1987.

\end{thebibliography}
					\end{document}